\documentclass[twocolumn,aps,prl,showpacs]{revtex4}
\usepackage{graphics,amsmath,graphicx}

\begin{document}
\title{Substrate limited electron dynamics in graphene}

\author{S. Fratini$^{1,2}$, F. Guinea$^2$}

\affiliation{$^1$ Institut N\'eel - CNRS \& Universit\'e Joseph Fourier,
BP 166, F-38042 Grenoble Cedex 9, France\\
$^2$  Instituto de Ciencia de Materiales de Madrid. CSIC, Sor Juana In\'es de
la Cruz 3, E-28049 Madrid, Spain}

\begin{abstract}
We study the effects of polarizable substrates such as SiO$_2$ and SiC on
the carrier dynamics in graphene. We find that 
the quasiparticle spectrum acquires 
a finite broadening  due to the long-range interaction with the
polar modes at the interface between graphene and the
substrate. This mechanism results in a density dependent electrical
resistivity, that exhibits a sharp increase around room
temperature, where it can  become the dominant 
limiting factor of electron transport. The effects  are weaker in doped bilayer
graphene, due to the more conventional parabolic band 
dispersion.
\end{abstract}
\date{\today}
\pacs{71.20.Tx  72.10.Di  73.63.-b  }
\maketitle

{\em Introduction.}
Remote phonon scattering is a known limiting factor of the electron
mobility in two-dimensional artificial structures of technological
interest, such as Si MOSFETs \cite{WM72,HV79,FNC01,Cetal04}. Due to the polar
nature of the gate dielectrics used in such devices, the carriers in
the conducting channel  couple electrostatically to the long-range
polarization field created at the conductor/dielectric interface.
The interaction occurs primarily
with optical phonon modes of high frequency ($50-200$ meV),
which do not affect transport
at low temperatures. However, it results in a sizable
degradation of the mobility at room temperature (i.e. the temperature
of device operation), and becomes the dominant scattering
mechanism in devices with high-$\kappa$ gate dielectrics.

The interaction with the phonons of the gate dielectric
has been recently investigated in another class of
electronic devices,  where the Si inversion layer
is replaced by a crystalline organic semiconductor \cite{Setal04,Hetal06}.
In organic FETs, due to the narrow-band nature
of the active material (typical bandwidths are less than $0.5$ eV),
the energy scale associated to the remote phonon interaction can become
comparable with the bandwidth, resulting in carrier self-localization. In
this case the effect is more dramatic than in Si MOSFETs:
the mobility is not only strongly reduced by the interactions, but it also
acquires an exponentially activated temperature dependence, due to the
hopping motion of the localized carriers.

Graphene --an atomically thin sheet of carbon atoms-- is
another system where the long-range coupling to the polar
modes of the substrate can have a sizable influence on carrier
dynamics.  Compared to the two-dimensional electron gas formed in Si
inversion layers,  the effect here is enhanced due to the
poor screening properties of the  quasiparticles close to the nodal 
points, which behave as Dirac fermions.   
Although most experimental studies of transport in graphene 
up to now 
have focused on the low temperature regime \cite{GN07,NGPNG07},
its possible use as a material basis for future electronic devices calls for
a deeper understanding of the room temperature
behavior \cite{HGNSB06,Tetal07,T2etal07,Metal07,private}. 

In this work we calculate the effects
of remote phonon scattering on the dynamics of electrons in
graphene \cite{Netal05b}. Unlike other scattering mechanisms such as
disorder \cite{NM07,AHGS07,Z07,SPG07,KG07}, or acoustic phonon
scattering \cite{HS07}, all the microscopic parameters of the remote
electron-phonon interaction are known \textit{a priori} from
independent measurements, so that  in principle its effects can be estimated
quantitatively.
We show that in epitaxial graphene grown on SiC, 
the polar phonon scattering has only a weak effect on the electron
mobility,   due to the weak polarizability of the substrate and  the
relatively high phonon frequencies associated to the hard Si-C bonds.
The effect is  much larger at graphene/SiO$_2$
interfaces where, at room temperature and at the typical densities
$n\gtrsim 10^{12}cm^{-2}$ attained in current experiments,
it gives rise to resistivities of the order of $100 \Omega$.
This suggests that, as in traditional Si based electronic devices,
remote scattering with the substrate phonons
can indeed constitute an important limiting factor of the mobility in
future graphene devices.

\paragraph{The model.}

The polar phonons of the substrate
induce an electrostatic potential which couples to the electron charge. 
This potential is smooth over
lengths comparable to the lattice constant,
due to the finite distance between the substrate and the
graphene layer (see below), which  allows us to neglect intervalley
scattering. In this case
the electron-phonon interaction term is given by $H_I=\sum_q M_q
\rho_q (b_q+b^+_{-q})$,  
where $\rho_q$ is the electron density operator 
and the operators
$b^+_q,b_q$ describe the phonon displacements. The interaction
matrix element can be written as $M_q^2=g e^{-2qz}/(qa)$
\cite{WM72,HV79,MA89,FNC01}, where $z$ is
the distance between the graphene layer and the substrate and $a=1.42$
\AA \
is the lattice spacing.  Neglecting the dielectric
response of the atomically thin graphene layer,  the
dimensionless coupling parameter $g$ is given by
\begin{equation}
  \label{eq:coupling}
  g=2 \pi \beta \frac{\hbar \omega_s}{\hbar v_F/a} \frac{e^2}{\hbar v_F}.
\end{equation}
Here $\beta=(\epsilon_s-\epsilon_\infty)/(\epsilon_s+1)/(\epsilon_\infty+1)$
is a combination of the known dielectric constants of the substrate,
$v_F=10^6$ cm/s the Fermi velocity
and $\omega_s$ the frequency of the interface phonon mode under
consideration.
We consider separately the case of epitaxial graphene
grown on SiC, and that of  graphene flakes  deposited on SiO$_2$ substrates.

In 6H-SiC there is a single surface phonon mode at
$\omega_s=116$meV \cite{NKM89}. The dielectric constants of the bulk
material are $\epsilon_s=9.7$, $\epsilon_\infty=6.5$, which yields
$\beta=0.040$ and  $g=1.7\cdot 10^{-2}$.
In the case where several surface modes are present, as in
SiO$_2$, each mode will couple through a partial factor $\beta_i$
weighted by its
oscillator strength (the sum of all $\beta_i$'s still given by
$\beta$ defined above). For example, in crystalline
SiO$_2$ ($\epsilon_s=3.9$, $\epsilon_\infty=2.4$),
there are two dominant surface modes at $\omega_{s1}=59$ meV and
$\omega_{s2}=155$ meV, with $\beta_1=0.025$ and $\beta_2=0.062$
respectively ($\beta=0.087$)  \cite{FNC01}, corresponding to
$g_1=5.4\cdot 10^{-3}$ and $g_2=3.5\cdot 10^{-2}$. These values are
enhanced by roughly $50\%$ in common SiO$_2$ glass, where
$\epsilon_\infty=2.1$. For completeness, we also give the values
for the high-$\kappa$ dielectric HfO$_2$ that has been used in
recent experiments on locally gated graphene \cite{OJEK07}:
$\epsilon_s=22$, $\epsilon_\infty=5$, $\beta=0.12$ and
$\omega_s\simeq 94 meV$, so that $g=4.2\cdot 10^{-2}$.

\paragraph{Scattering rate.}
The quasiparticle scattering rate
arising from the remote electron-phonon interaction described above
can be calculated as 
 \begin{eqnarray}
  \label{eq:tau}
  \Gamma(\omega)&=&\frac{\pi}{2}\sum_{\bf q} A_{{\bf p},{\bf q}} 
M_q^2 \left\lbrace
  \delta(\epsilon_{\bf p+q}-|\omega+\omega_s| ) [n_B+n_F^+]
  \right. \nonumber \\
& & + \left.
   \delta(\epsilon_{\bf p+q}-|\omega-\omega_s|)
  [n_B+1-n_F^-]
\right\rbrace,
\end{eqnarray}
where the band dispersion is $\omega=\pm \epsilon_{\bf p}=\pm v_F p$,
$n_B$ is  the Bose  distribution for phonons
of energy $\omega_s$  and $n_F^+, n_F^-$ are the Fermi
functions  for electrons at $\omega\pm \omega_s$.
The factor $A_{{\bf p},{\bf q}}=[1+s\cos (\phi_{\bf p+q}-\phi_{\bf p})]/2$ is the spinor overlap
for intraband $(s=1)$ and interband $(s=-1)$ scattering respectively,
where $\phi_{\bf p}$ defines the direction of ${\bf p}$. 
To a first approximation,  dynamic screening from the conduction
electrons can be accounted for by replacing $1/q\to 1/(q+q_s)$ in
the interaction matrix element $M_q^2$, where
$q_s=4e^2k_F/\hbar v_F$ is the Thomas-Fermi screening
length \cite{WSSG07}. We set the graphene/substrate distance to
the typical value $z=4$\AA.
Eq. (\ref{eq:tau}) can be directly generalized to bilayer graphene, by
considering a parabolic band dispersion $\epsilon_p=v_F^2
p^2/t_\perp$, with  $t_\perp\sim 0.35 eV$  the
interlayer hopping parameter. In that case the screening  wavevector
becomes independent of band filling, $q_s=4t_\perp /\hbar v_F$, and
the spinor overlap changes to $[1+s\cos 2(\phi_{\bf p+q}-\phi_{\bf p})]/2$.
We also take a slightly larger separation $z\simeq 6$\AA, 
corresponding to the average distance of the bilayer to
the interface, although in principle the electron-phonon 
coupling differentiates between the two graphene sheets. 
\begin{figure}
  \centering
\includegraphics[width=70mm]{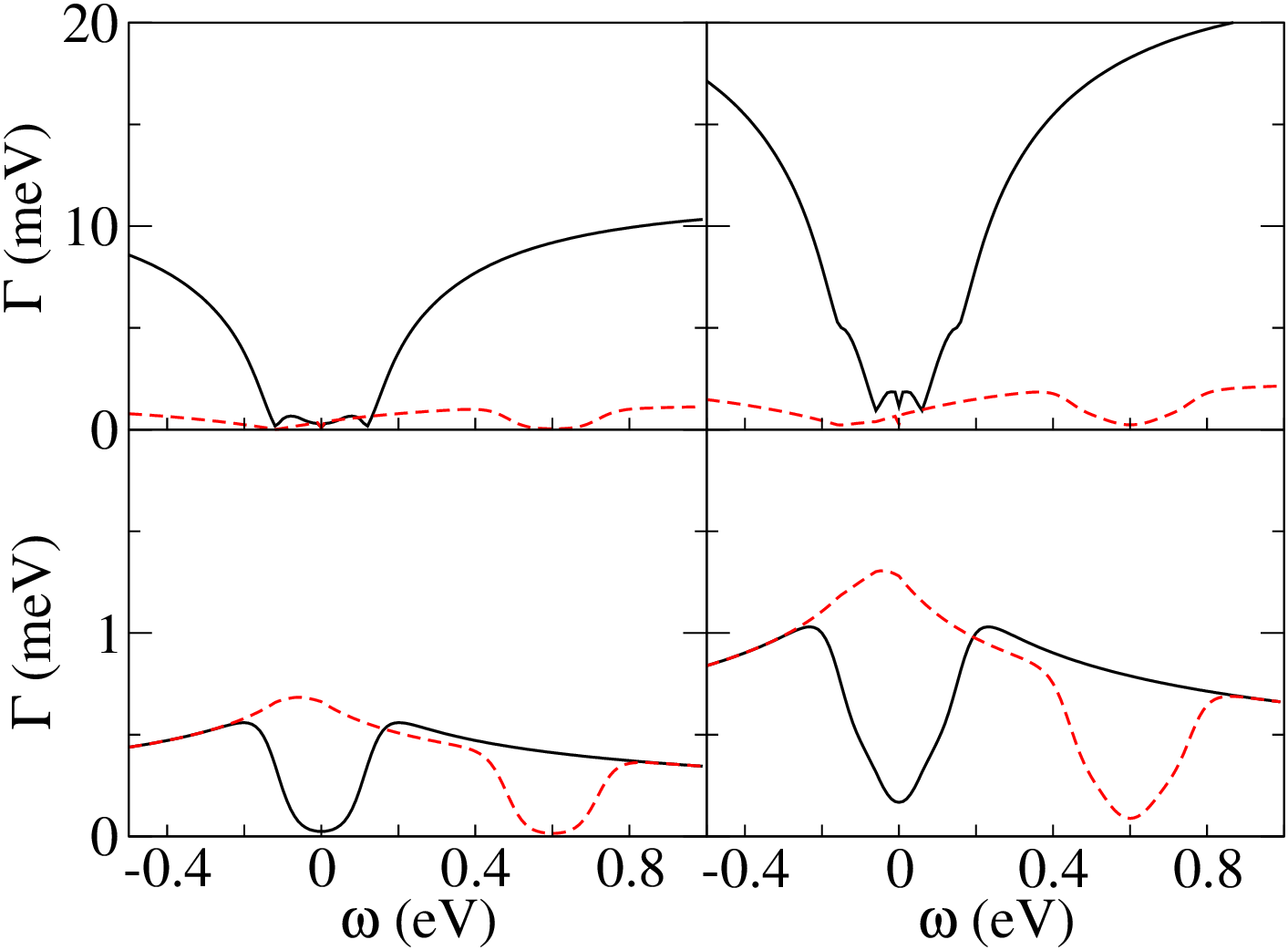}
\caption{Quasiparticle's scattering rate due to remote phonon
  interaction  as function of energy for:
  single-layer (top) and  bilayer (bottom) graphene, with the parameters
appropriate to an SiC (left) and a SiO$_2$ (right) substrate
respectively. The full and dotted lines correspond
respectively to undoped, $E_F = 0$, and heavily doped, $E_F = 0.6$eV
cases.  Notice the different axis scale in single-layer (top)
and bilayer graphene (bottom).}
\label{fig:scattering}
\end{figure}

In Fig.\ref{fig:scattering} we show plots of the scattering rate
$\Gamma$  corresponding to single-layer
and bilayer graphene  on the two different substrates
SiC and SiO$_2$, for both undoped  ($E_F=0$) and heavily
doped ($E_F=0.6$eV) graphene.
In all cases the scattering rate is suppressed
in a window $\pm \omega_s$ around the
Fermi energy (it actually vanishes at $T=0$,
because in that case phonon exchange is
forbidden in all this range). 
The absolute value of $\Gamma$ at the chemical
potential, that will enter in the calculation of the d.c. conductivity,  
is proportional to both the coupling strength $g$ and
the phonon thermal population, i.e.
$\Gamma \propto g e^{-\omega_s/T}$. 
Note that the anomalously large scattering rates obtained for 
undoped single-layer graphene (full lines in the upper panels of Fig. 1) 
compared to all the other cases are related to the poor screening 
properties of Dirac electrons close to
the neutrality point.

The present results for the quasiparticle lifetimes from remote phonon
scattering can  be compared
with analogous estimates for the interaction with the intrinsic in-plane
vibrations of the graphene sheet \cite{G81}.
The effect of the substrate appears to be slightly weaker than the intrinsic
electron-phonon scattering in the case of
epitaxial graphene grown on SiC, while the two mechanisms are of
comparable magnitude at graphene/SiO$_2$ interfaces. 
Nevertheless, we will show 
below that due to the smaller  oscillation frequencies involved,
remote phonons  have a much stronger influence 
on electronic transport than the intrinsic phonons of graphene.

\paragraph{Conductivity.}
We calculate the electrical conductivity through the Boltzmann equation.
Taking into account  spin and valley degeneracy, the d.c. conductivity of
single-layer graphene can be expressed as:
\begin{equation}
  \label{eq:sig-bol}
  \sigma=\frac{e^2}{h} \int d \omega |\omega| \Gamma_{tr}^{-1}(\omega)
\left(- \frac{dn_F}{d\omega}\right).
\end{equation}
Here $\Gamma_{tr}(\omega)$ is the transport scattering rate, defined
as in Eq.(\ref{eq:tau}), with the additional angular factor $[1-s\cos
(\phi_{\bf p+q}-\phi_{\bf p})]$  
in the integrand which favors large angle scattering
events  \cite{M90}. 
Note that this approximation is strictly valid only in the elastic
limit $\omega_0\to 0$ (where  $\Gamma_{tr}\approx \Gamma/2$) while it 
slightly underestimates the scattering rate in the inelastic
regime  $\omega\lesssim \omega_0$.
In the case of bilayer graphene, the integral in Eq. (\ref{eq:sig-bol})
has an additional factor $2\omega/t_\perp$ arising
from the parabolic band dispersion.

We now present results for graphene on SiO$_2$ substrates, where most
experimental transport measurements have been performed (the predicted
effect at room temperature is one order of magnitude smaller in
the case of SiC). From
Eq. (\ref{eq:sig-bol}), since the derivative of the Fermi function
selects contributions close to the Fermi energy,
we see that the conductivity is inversely
proportional to the scattering rate  at $\omega\simeq E_F$, which in
turn scales as $g e^{-\omega_{s}/T}$ (see above). 
Since the coupling constant $g$ in Eq. (\ref{eq:coupling})
is directly proportional to the phonon frequency $\omega_{s}$, we see that
the scattering rate at a given temperature $T$ is a non-monotonic
function of the phonon frequency:  it is
maximum at $\omega_{s}=T$, and  decreases exponentially for larger
phonon frequencies. As a result, the phonons that mostly affect
transport will be the ones with the lowest frequencies (closest to $T$).
In particular, in the case of a SiO$_2$ substrate,
the dominant phonon is the one at $\omega_{s1}=59$meV, 
even though it has a weaker coupling
strength (due to the exponential temperature dependence, scattering
from the phonon at $\omega_{s2}=155$meV is mostly inactive at room
temperature). For the same reason, in-plane optical phonons whose
characteristic energy scale is $\sim 200$meV should give a
negligible contribution to the resistivity at room temperature.

\begin{figure}
  \centering
\vspace{4.6cm}
  \includegraphics[width=67mm,height=42mm]{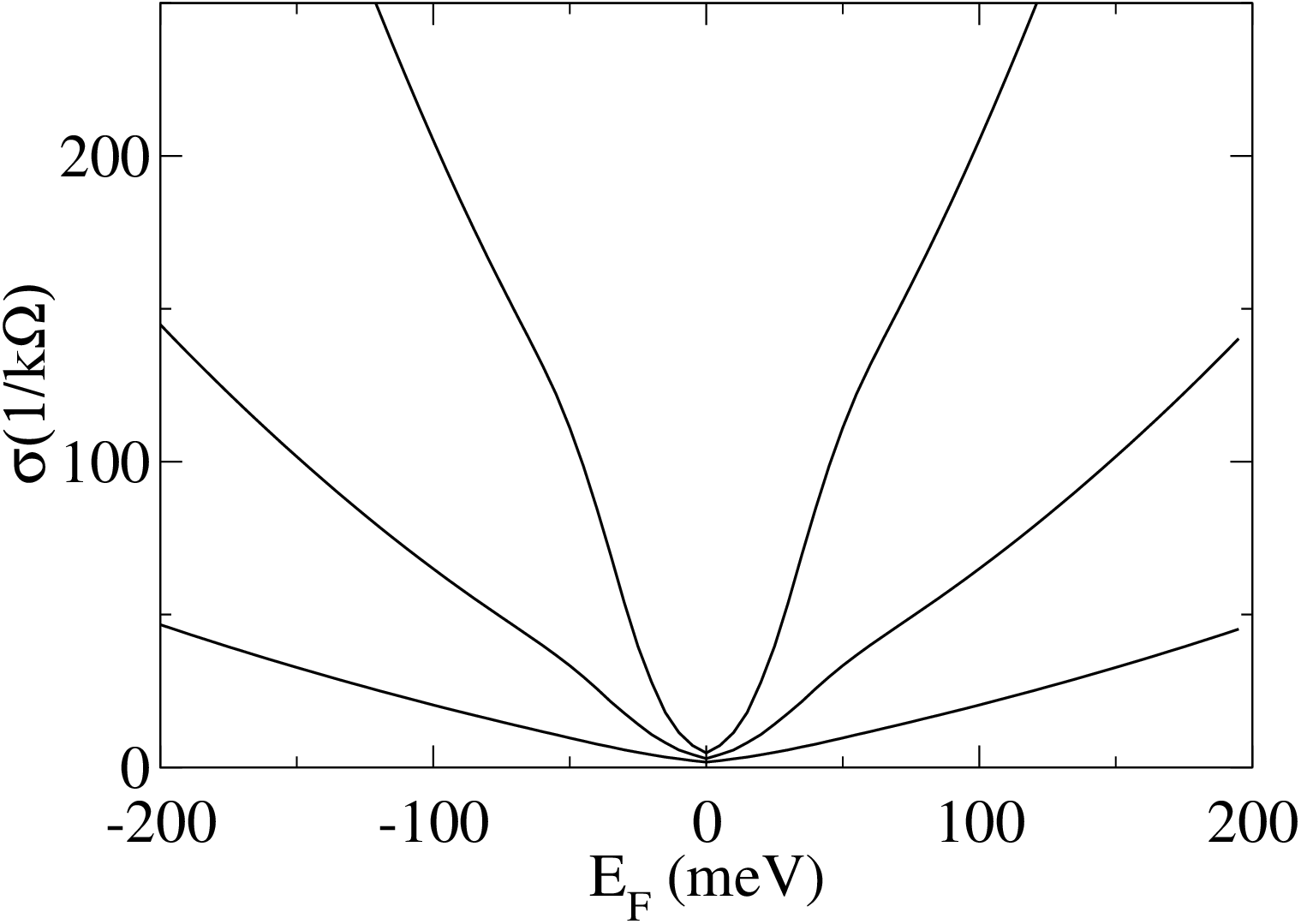}\\
\vspace{-8.6cm}
  \includegraphics[width=65mm,height=42mm]{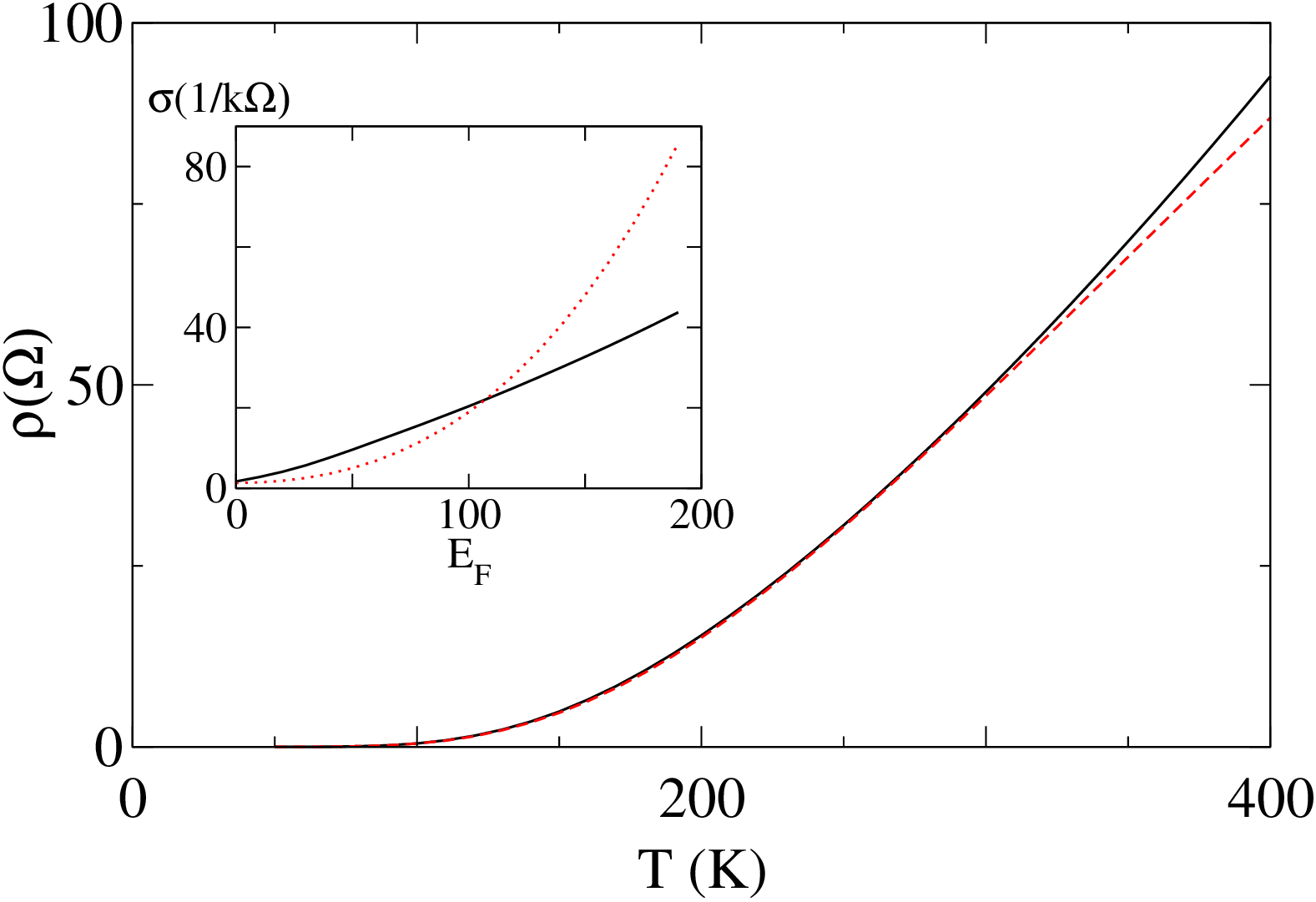}  \\
\vspace{4.cm}
\caption{a) Temperature dependence of the resistivity for $E_F =
    100$ meV (the dashed line corresponds to $\rho=\rho_0
e^{-\omega_{s1}/T}$ with $\rho_0=500 \Omega$).
b) Conductivity vs. chemical potential for T = 150, 200
and 300K  (from top to bottom). The inset of a) shows the comparison with
bilayer graphene (dotted line) at $T=300K$.}
  \label{fig:transport}
\end{figure}

\paragraph{Results.} 
In Fig. \ref{fig:transport}a we show the result of $\rho$ vs $T$ at
$E_F=100$meV (corresponding to a density $n=0.7 \cdot 10^{12}cm^{-2}$)
taking a finite distance $z=4$\AA \
between the substrate and the graphene layer
and a Thomas-Fermi screening from the graphene electrons as
described above \cite{screening}. As anticipated from the preceding discussion, 
the resistivity exhibits a characteristic exponential behavior. 
Note that  unlike the short-range 
scattering from in-plane phonons, the present result
has a sizable density dependence due to the long-range
nature of the interaction. 
This is illustrated in Fig.\ref{fig:transport}b, where we plot the
dependence
of $\sigma$ on $E_F$ at fixed  $T=300,200,150$K (from bottom to top).
In the explored density range, the conductivity has an overall linear
dependence on $E_F$ away from the Dirac point, which is
intermediate between the cases of short range scatterers ($\sigma \sim
const $) and charged impurities ($\sigma \propto E_F^2$)\cite{NM07}. This
behavior can be understood from standard dimensional arguments,
observing that the scattering rate of Eq.(\ref{eq:tau}) scales as
$M^2_{q\simeq k_F}\propto k_F^{-1}$  
times the density of states $\propto k_F$, which yields  
$\Gamma\sim const$. The result $\sigma \propto |E_F|$ then follows directly
from Eq. (\ref{eq:sig-bol}).

Superimposed to the  smooth density dependence analyzed above,  
a slight ``kink'' appears in the electrical
characteristic at $E_F\simeq \omega_{s1}$
that testifies the interaction with the sharp interface phonon mode. 
The anomaly becomes more pronounced at low temperatures
owing to the sharp Fermi function in the Kubo formula. There,
however,  the absolute 
value of the scattering rate becomes extremely small, and it is likely to be
completely hidden by the presence of other sources of scattering.

In the inset of Fig. \ref{fig:transport}b we compare the conductivity
for single-layer and bilayer graphene at $T=300K$. The
different density of states in bilayer graphene 
leads to a stronger energy dependence. Accordingly, bilayer graphene 
should be a better conductor at large bias. With the present
choice of parameters, the crossing occurs at $E_F\sim
0.1eV$.

\paragraph{Finite distance cutoff.}
The simple scaling argument presented above, leading to a linear
dependence  $\sigma \propto |E_F|$ for single layer graphene, breaks
down when $2 k_F z \sim 1$. At larger densities, 
the short range cut-off associated to
the distance $z$ between the carriers and the interface must be
taken into account explicitly. As a result, extra powers of $k_F
z$ arise in the scattering rate, due to the suppression of 
the angular factors $[1\pm \cos (\phi_{\bf p+q}-\phi_{\bf p})]$
at large scattering angles, as
well as the reduced available density of states. All together, these
factors  eventually lead to a $\sigma \propto E_F^4$ dependence
($\Gamma \propto |E_F|^3$). 
In Fig.3 we show plots of $\sigma$ vs $E_F$ at $T=300K$ for increasing 
values of $z$. While for $z=4$ \AA \ the behavior is approximately 
linear in the whole energy range, a clear upturn becomes visible 
at larger separations, reflecting the expected crossover  at 
$|E_F|z/v_F\sim 1/2$. Notably, this phenomenon can lead to an
apparent behavior $\sigma \propto E_F^2\propto n$ \cite{private}, 
which is best seen in the inset of Fig.3.

\begin{figure}
  \centering
  \includegraphics[width=65mm,height=45mm]{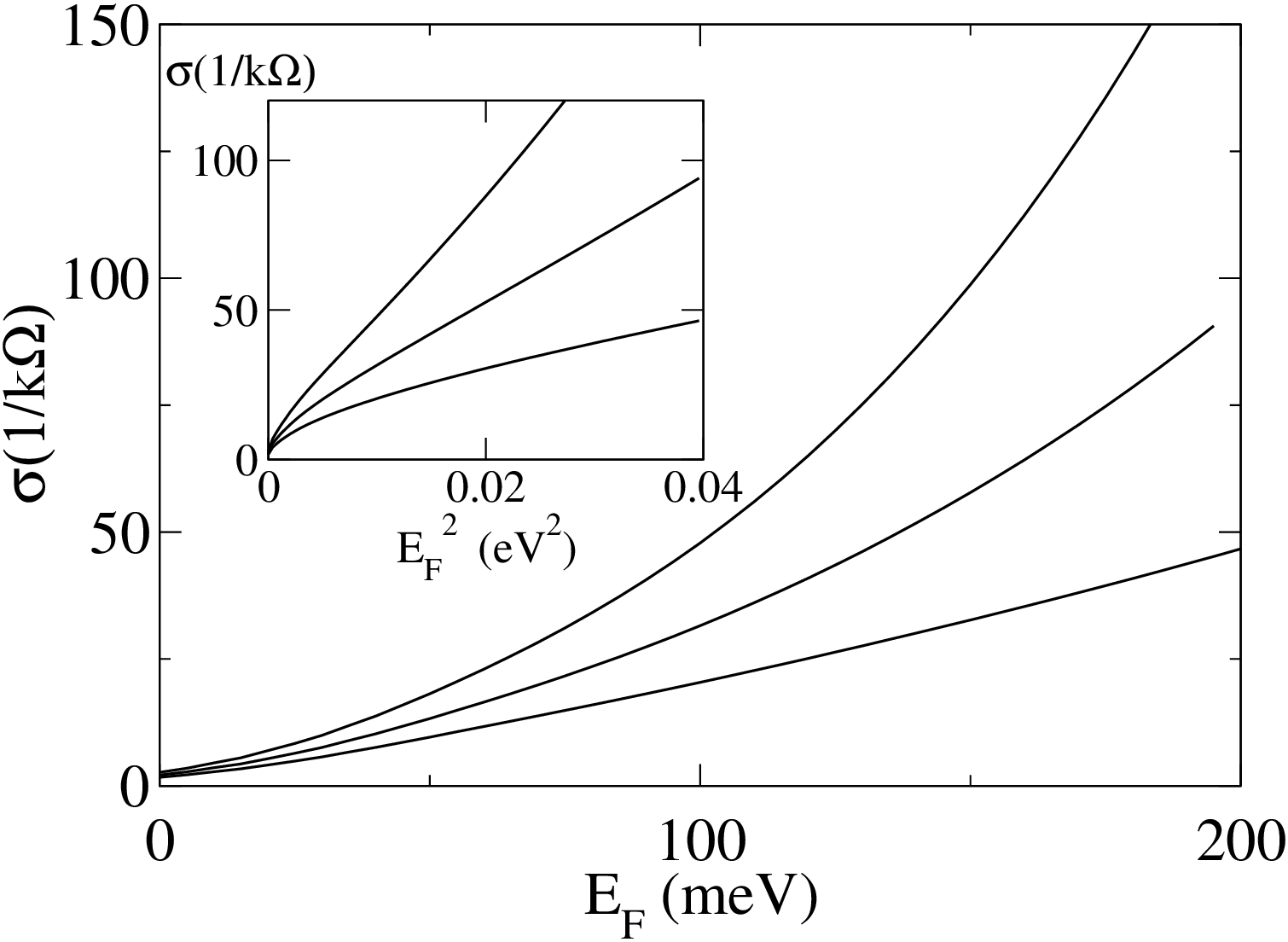}
\caption{Same as Fig 2.b, for different values of the distance
  $z$. From bottom to top, $z=4,
  12$ and $20$ \AA. A crossover to a super-linear behavior is apparent
  at large $z$ and large $E_F$. In the inset, the same data are
  plotted as a function of $E_F^2\propto n$.}
\end{figure}

\paragraph{Discussion.}
Although the transport properties of graphene are roughly temperature
independent  below $T=100$K, a sharp increase of the resistivity of the
order of $\Delta \rho\sim 100 \Omega$ around room temperature
has been reported in field-effect doped samples at $n\gtrsim 10^{12}$cm$^{-2}$
\cite{HGNSB06,Tetal07,T2etal07,Metal07,private}. 
This increase is unlikely to  be due to the effects of disorder,
which are expected to depend smoothly, if at all, on the
temperature. As was mentioned above, also the in-plane optical phonons of
graphene can be ruled out due to their high oscillation frequencies.
That in-plane scattering is not the main cause of the observed behavior
is further supported by the observation
 that the  temperature dependent contribution to the resistivity is
 not independent of density \cite{Metal07,private},  pointing to a
 long-range scattering mechanism such as the one studied here.  
According to the present results, a density and temperature dependent 
contribution to the resistivity 
is compatible with the mechanism of remote phonon scattering.

\paragraph{Conclusions.}
We have estimated the quasiparticle scattering rates induced by the polar
modes of SiO$_2$ and SiC substrates on graphene.
Our results show that, while remote phonon scattering
has negligible effects on transport at low temperatures, 
it becomes relevant at room temperature, where
it could constitute the dominant limiting factor
of the electron mobility in sufficiently clean samples, especially for
SiO$_2$ substrates. According to the present scenario, 
the maximum conductivity that can be achieved in a given sample
depends crucially on the dielectric properties of the gate
insulator, so that the use of non-polar substrates
should be favored to optimize the transport  properties of graphene.
We also find that the effects 
are reduced in doped
bilayer graphene, which makes it more suitable for device
applications at room temperature. On the opposite, larger effects
can be expected when more polarizable substrates  
are used as gate dielectrics.

\paragraph{Acknowledgements.}
This work was supported by MEC (Spain) through grant
FIS2005-05478-C02-01, the Comunidad de Madrid, through the program
CITECNOMIK, CM2006-S-0505-ESP-0337, and the European Union Contract
12881 (NEST). 
The authors thank M.S. Fuhrer for stimulating correspondence and T. Stauber 
for useful discussions.
\bibliography{substrate-sub}

\begin{thebibliography}{27}
\bibitem[{scr()}]{screening}
\bibinfo{note}{This approximation assumes that the electrons respond
  instantaneously to the lattice vibrations, and therefore constitutes a lower
  bound to the scattering rate. An upper bound is obtained by plainly
  neglecting screening which in the regime of interest here leads to
  resistivities $\approx 4$ times larger than the ones shown.}
\end{thebibliography}
\end{document}